\documentclass[journal]{IEEEtran}
\usepackage{graphicx}
\usepackage{babel}
\usepackage{amsmath,bbm}
\usepackage{amssymb}
\usepackage{color}
\usepackage{cite}

\usepackage{mathtools}

\ifCLASSINFOpdf

\else

\fi

\newcommand{\Tr}{\textrm{Tr}}
\newcommand{\Var}{\textrm{Var}}

\newcommand{\eqnref}[1]{Eq.~(\ref{#1})}

\newcommand{\win}{\omega_{\text{in}}}
\newcommand{\wout}{\omega_{\text{out}}}

\newcommand{\Mout}{M_\text{out}}
\newcommand{\Mbout}{\bar{M}_{\text{out}}}
\newcommand{\wbout}{\bar{w}_{\text{out}}}
\newcommand{\wbin}{\bar{w}_{\text{in}}}
\newcommand{\nbar}{\bar{n}}

\begin{document}

\title{Quantum Limits on the Capacity of Multispan Links with Phase-sensitive Amplification}

\author{Karol~\L{}ukanowski, Konrad~Banaszek and~Marcin~Jarzyna%
\thanks{The authors are with the Centre for Quantum Optical Technologies, Centre of New Technologies, University of Warsaw, Banacha 2c, 02-097 Warszawa, Poland. K.\L{}. and K. B. are also with the Faculty of Physics, University of Warsaw, Pasteura 5, 02-093 Warszawa, Poland}%
}

\maketitle

\begin{abstract}
Long-distance fiber communication stands as a cornerstone of modern technology. One of the underlying principles, preventing signal levels from diminishing below the detectability threshold, is optical amplification. In particular, phase-sensitive amplifiers offer a promising solution as ideally they do not introduce any excess additive noise. Since such devices in principle operate at the quantum noise level, a natural question is whether one can further improve the capacity of amplified links using principles of quantum mechanics as it offers a much broader scope of signal modulations and detection schemes. We derive ultimate limits determined by the laws of quantum mechanics on the capacity of multispan links with phase sensitive amplification. We show that the quantum advantage over the standard approach based on optical quadrature detection is small and vanishes for long links.
\end{abstract}

\IEEEpeerreviewmaketitle

The technological demand for the constantly raising amount of information exchanged between different entities puts extensive pressure on the increase of the communication rates of optical fiber links \cite{Essiambre2012}. One of the main factors that limits the performance of a communication link is the reduction of the signal-to-noise ratio (SNR) due to the presence of losses in a fiber cable or other parts of an optical link. In order to overcome this issue, one may investigate various techniques such as changing the fiber structure or using different modulation formats \cite{Essiambre2012, Kikuchi2016, Kahn2004, Winzer2012}. The primary way, however, is to incorporate signal amplification, by which one can restore signal power to a desired level. Standard phase-insensitive amplifiers allow to bring back signal level at the cost of introducing additional noise. Importantly, this noise cannot be reduced below a certain value because of fundamental quantum mechanical effects \cite{Caves1982}. This phenomenon causes a decrease of SNR with link length, since the noise introduced by each amplifier in the cable is amplified by subsequent ones. The overall effect is that one can indeed vastly improve communication rates with conventional phase-insensitive amplification, however, the rate still inevitably decreases with the link length \cite{Yariv1990, Antonelli2014, Jarzyna2019a}.

A more sophisticated method of signal restoration is phase-sensitive amplification \cite{Kakande2011, Umeki2013, Olsson2018}. A phase-sensitive amplifier (PSA) in general can amplify one of the signal quadratures while simultaneously reducing the other one \cite{Caves1982}. At first sight, this is detrimental to the capacity, since one is able to efficiently transmit information encoded only in a single (amplified) quadrature of light whereas in the previous case both quadratures could carry the information. The advantage, however, is that PSAs are in principle noiseless devices, i.e., they do not introduce any additive noise. It is therefore expected that for large distances, when one would like to use a significant number of amplifiers, they can lead to an improved SNR as compared with the phase-insensitive scenario.  

On the other hand, one may go beyond the standard picture of classical information theory and instead inspect a broader range of strategies allowed by the laws of quantum mechanics. In particular, novel detection schemes \cite{Burenkov2021}, such as the Dolinar receiver \cite{Dolinar1973, Banaszek2020} or various collective receivers \cite{Guha2011} make it possible in certain scenarios to surpass classical capacity limits imposed by the Shannon-Hartley bound and attain the fundamental quantum capacity limit given by the Gordon-Holevo expression \cite{Gordon1962, Holevo1973, Westmoreland1997, Holevo1998}. Crucially, for such more general strategies one has to abandon SNR as a figure of merit and instead consider signal and noise separately. Another possibility for improvement is to use a non-standard modulation format that utilizes such quantum features of light as squeezing which in principle seems to be beneficial for phase sensitive channels \cite{Schafer2016}.

In this article we analyze quantum capacity limits of multispan links in the presence of PSA. We identify two important regimes of amplification, depending on whether just the signal or the total power are restored to their initial values. We show that in the large distance limit in the former regime one obtains an exponential gain in the capacity with respect to the unamplified scenario. On the other hand, for total power restoration the capacity maintains its exponential decay known from the pure loss channel instance but with an advantage in the form of an improved exponent when compared to both unamplified and phase-insensitive amplification cases. In the range of more typical distances, up to a few thousand kilometers, depending on the signal strength, one observes an exponential advantage in SNR and the capacities are approximately equal in both approaches. Importantly, the enhancement attainable by the most general quantum strategies over the standard Shannon information bound is present only for low distances and becomes negligible for large fiber lengths, meaning that quadrature detection is a nearly optimal detection strategy.

\section{PSA fiber link}

\begin{figure}
	\centering
	\includegraphics[width=1.0\columnwidth]{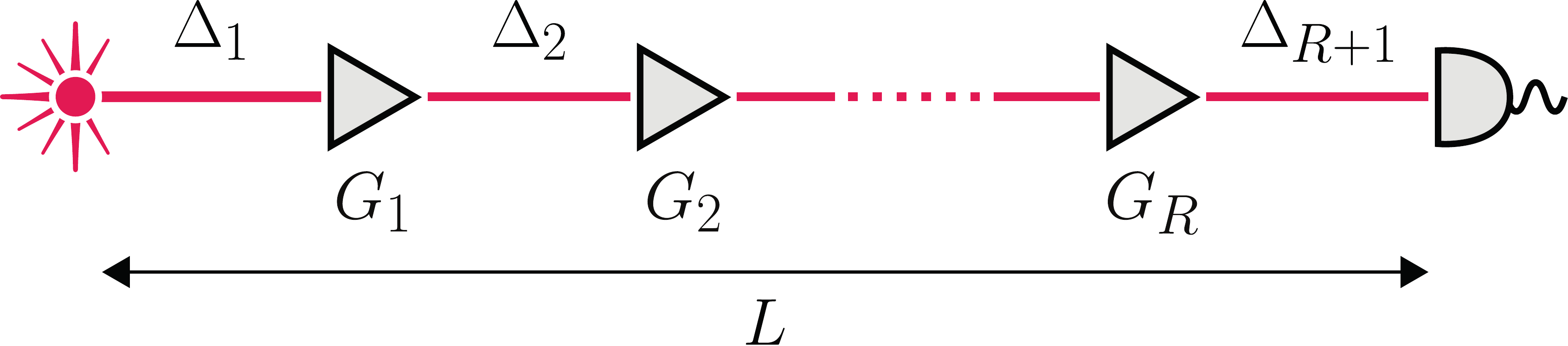}
	\caption{Scheme of a lossy channel of length $L$ with $R$ amplifiers with gains $G_r$ separated by distances $\Delta_r = l_r - l_{r-1}$ from each other.}
\label{fig:scheme}
\end{figure}

A basic model of a PSA fiber link, presented schematically in Fig.~\ref{fig:scheme}, consists of a standard lossy channel characterized by an attenuation constant $\alpha$, with total length $L$ and with $R$ PSAs inserted at regeneration nodes. The i-th amplifier is specified by a gain $G_i$ and is located at a distance $l_i$ from the input, $i=1,2,\dots R$. An optical field is characterized by two orthogonal quadratures, denoted by $x^Q$ and $x^I$. These quadratures can be further decomposed as
\begin{equation}\label{eq:quad}
    x^{Q/I}=x^{Q/I}_S+x^{Q/I}_N,
\end{equation}
where $x^{Q/I}_S$ represents the contribution of the signal and $x^{Q/I}_N$ describes the noise. Since noise and signal are not correlated the quadratures variances can be similarly decomposed
\begin{equation}\label{eq:var}
    \Var[x^{Q/I}]=S^{Q/I}+N^{Q/I},
\end{equation}
where $S^{Q/I}=\Var[x^{Q/I}_S]$ denotes the signal power in each quadrature and $N^{Q/I}=\Var[x^{Q/I}_N]$ are the corresponding noise powers. Assuming that no external additive noise is introduced at any point of the link, the quadratures and their variances after the $i$-th amplifier are equal to\vspace{2\jot}
\begin{IEEEeqnarray}{rClrCl}
x^{Q}_i & = & \mathrlap{x^{Q}_{i-1}\sqrt{G_i \tau _i}+\sqrt{1-\tau_i} \; x^Q_{N,i},}\label{eq:xq}\\
x^{I}_i & = & \mathrlap{x^{I}_{i-1} \sqrt{\frac{\tau_i}{G_i}} + \sqrt{1-\tau_i} \; x^I_{N,i},}\label{eq:xi}\\
S^Q_i & = & G_i\tau_i S^{Q}_{i-1}, & \quad
N^Q_i & = & G_i\left(\tau_i N^Q_{i-1} + \frac{1-\tau_i}{2}\right), \label{eq:var_q}\\
S^{I}_{i} & = & \frac{\tau_i}{G_i}S^{I}_{i-1}, & \quad
N^{I}_{i} & = & \frac{1}{G_i} \left(\tau_iN^I_{i-1} + \frac{1-\tau_i}{2}\right),\vspace{2\jot}
 \label{eq:var_p}
\end{IEEEeqnarray}
where $\tau_i=e^{-\alpha (l_i-l_{i-1})}$ and $x^{Q/I}_{N,i}$ denote respectively transmission and noise variables contributed by the i-th span. Importantly, since amplifiers are phase sensitive, only one quadrature is amplified; the $Q$ quadrature for $G_i\geq 1$ and the $I$ quadrature for $G_i\leq 1$. Note also that we assumed in \eqnref{eq:xq}-(\ref{eq:var_p}) that the amplification process is aligned with the quadrature basis of the signal, i.e., it does not introduce correlations between $x^Q$ and $x^I$. This requires some kind of phase stabilization between the source and the PSA.

In the absence of losses, $\tau_i=1$, PSA just rescales quadratures and their variances correspondingly. This means that it can amplify a particular quadrature without introducing any additive noise. Crucially, however, when the channel is nonideal, $\tau_i<1$, there appears a second term in the expression for $N^{Q/I}$ in \eqnref{eq:var_q} and \eqnref{eq:var_p} which increases the noise. This term originates from the amplification of the vacuum fluctuations that entered the signal during losses in the $i$th span in \eqnref{eq:xq} and \eqnref{eq:xi}. Note that these fluctuations are then amplified by each remaining PSA in the link which can make their contribution substantial at the output. Therefore, even though PSA is a noiseless process, due to the interaction between amplification and losses in the link, one cannot completely eliminate this phase sensitive noise.

In the quantum mechanical description, the optical field is characterized by quantum states of light and the quadrature random variables are promoted to operators \cite{Banaszek2020}. A particularly important class of states, which we will consider here, are Gaussian quantum states which are fully characterized by the first and second quadrature operators moments. They include a wide variety of practically relevant states of light such as coherent states, which are quantum analogues of classical electromagnetic waves, and squeezed states. Crucially, both phase-sensitive and insensitive amplification as well as losses are examples of Gaussian quantum operations which preserve the Gaussian features of quantum states, meaning that if the input state is Gaussian the output of the link would be Gaussian as well \cite{Weedbrook2011}.

Assuming one considers only Gaussian states, the classical description in \eqnref{eq:quad} and \eqnref{eq:var} can be also used to describe the quantum signal since one can take $x_S^{Q/I}$ as the first moment of the state and $x_N^{Q/I}$ as a value of a Gaussian random variable with distribution $x_N^{Q/I}\sim\mathcal{N}(0,V^{Q/I})$, where $V^{Q/I}$ denotes the fundamental quantum noise level corresponding to each quadrature. For coherent states $V^{Q/I}=1/2$ is the typical shot noise limit present in quadrature detection, but in general the noise just has to satisfy $V^QV^I\geq 1/4$ due to the Heisenberg uncertainty principle. One can therefore fully characterize the state of the received light pulses by looking at the quadrature values and their variances at the channel output which are equal to
\begin{IEEEeqnarray}{rClrCl}\label{eq:snr1}
    x^Q_{\textrm{out}}&=&x^Q_{\textrm{in}}\sqrt{\tau_{\textrm{tot}} G_{\textrm{tot}}},&\qquad x^I_{\textrm{out}}&=&x^I_{\textrm{in}} \sqrt{\frac{\tau_{\textrm{tot}}}{G_{\textrm{tot}}}},\\
    S^Q_{\textrm{out}}&=&S^Q_{\textrm{in}}\tau_{\textrm{tot}} G_{\textrm{tot}},&\qquad S^I_{\textrm{out}}&=&S^I_{\textrm{in}}\frac{\tau_{\textrm{tot}}}{G_{\textrm{tot}}}, \\
    N^Q_{\textrm{out}}&=&\mathrlap{\tau_{\textrm{tot}} G_{\textrm{tot}}N^Q_{\textrm{in}}+\frac{\tau_{R+1}}{2}\sum_{i=1}^R G_i(1-\tau_i) \prod_{j=i+1}^R G_j\tau_j}\nonumber\\
    &&{}+\frac{1-\tau_{R+1}}{2},
    \\
    N^I_{\textrm{out}}&=&\mathrlap{\frac{\tau_{\textrm{tot}}}{G_{\textrm{tot}}}N^I_{\textrm{in}}+\frac{\tau_{R+1}}{2}\sum_{i=1}^R \frac{1-\tau_i}{G_i}\prod_{j=i+1}^R \frac{\tau_j}{G_j}}\nonumber\\&&{}+\frac{1-\tau_{R+1}}{2},
    \label{eq:snr2}
\end{IEEEeqnarray}
where $\tau_{\textrm{tot}}=\prod_{i=1}^{R+1} \tau_i$, $G_{\textrm{tot}}=\prod_{i=1}^R G_i$, $x^{Q/I}_{\textrm{in}}$, $S^{Q/I}_{\textrm{in}}$ and $N^{Q/I}_{\textrm{in}}$ are the initial values of quadratures and signal and noise powers.

\section{Information theory}

In the standard information theory picture a general memoryless communication channel is characterized by a conditional probability distribution $p(y|x)$ which describes the statistical dependence of output symbols $y$ on the input ones $x$. The sender uses symbols $x$ with some prior probability distribution $p(x)$. In such a picture the communication rate is bounded by the mutual information
\begin{equation}\label{eq:mut_inf}
    I(X,Y)=H(Y)-H(Y|X),
\end{equation}
where $H(Y)=-\sum_y p(y)\log_2 p(y)$ and $H(Y|X)=-\sum_{x,y}p(x)p(y|x)\log_2 p(y|x)$ are the output and conditional Shannon entropies. One can optimize mutual information over the input probability distribution $p(x)$ in order to get the best performance and obtain the channel capacity
\begin{equation}\label{eq:cap_cl}
    C=\max_{p(x)}I(X,Y),
\end{equation}
which specifies the best achievable rate for a given information-theoretic channel.

In optical communication it is customary to impose some form of constraint on the input modulation, otherwise the capacity may become infinite. Typically, it is the average energy of the signal that cannot exceed some given value. This constraint can be expressed in terms of the average number of photons in the signal $\bar{n}$ and reads
\begin{equation}
\label{eq:input-signal-energy}
    S^Q_{\textrm{in}}+S^I_{\textrm{in}}=2\bar{n}.
\end{equation}
Under the above constraint and assuming heterodyne detection performed at the output one can derive the well known Shannon-Hartley bound
\begin{equation}\label{eq:double}
    C_{\textrm{S}2}=\frac{1}{2}\log_2\left(1+\textrm{SNR}^Q\right)+\frac{1}{2}\log_2\left(1+\textrm{SNR}^I\right),
\end{equation}
where $\textrm{SNR}^{Q/I}=S_{\textrm{out}}^{Q/I}/N_{\textrm{out}}^{Q/I}$ are the SNRs corresponding to $Q$ and $I$ quadratures respectively. PSA amplifies the signal in just one direction of the optical phase space while it reduces it in the orthogonal one. Therefore, for large distances, when the total accumulated losses are considerable, the PSA would reduce SNR in one direction in \eqnref{eq:double} to low values. Thus, it is beneficial to spend all energy on modulating just a single quadrature and perform homodyne detection. The resulting capacity is equal to
\begin{equation}\label{eq:single}
    C_{\textrm{S}1}=\frac{1}{2}\log_2\left(1+\textrm{SNR}^Q\right),
\end{equation}
where we decided to amplify the $Q$ quadrature and $S^Q_{\textrm{in}}=2\bar{n}$. Importantly, the capacity in the above expressions \eqnref{eq:double} and \eqnref{eq:single} is given solely by the signal-to-noise ratio (SNR) in respective quadratures.

The physical channel over which the information is transmitted at the fundamental level is characterized by the laws of quantum mechanics. In particular, the information about symbols $x$ is encoded in quantum states $\rho_x$ of physical information carriers, which in case of optical communication are photons. These states then undergo an evolution described by a quantum channel $\Lambda$ and are detected by the receiver using a measurement described by a positive operator valued measure (POVM) $\Pi_y$. The classical information theoretic channel can be then reconstructed using the Born rule $p(y|x)=\Tr[\Lambda(\rho_x)\Pi_y]$ which allows to evaluate both mutual information and channel capacity through \eqnref{eq:mut_inf} and \eqnref{eq:cap_cl} respectively. 

Crucially, in the quantum mechanical description one explicitly includes measurement and quantum states of the signal, meaning there are more degrees of freedom which can be used to boost information transfer rate of a physical channel $\Lambda$. In particular, optimization of the mutual information over measurements results in the Holevo bound \cite{Holevo1973}
\begin{equation}\label{eq:holevo}
    C_{\textrm{GH}}=S\left[\sum_xp(x)\Lambda(\rho_x)\right]-\sum_xp(x)S\left[\Lambda(\rho_x)\right],
\end{equation}
where $S(\rho)=-\Tr(\rho\log_2\rho)$ is the von Neumann entropy of a state $\rho$. The Holevo bound is in principle saturable but in general it requires exotic collective measurements performed on a very large number of channel outputs.

In order to find the upper bound on the capacity attainable for the general quantum measurements and states, one needs to evaluate \eqnref{eq:holevo} and optimize it over all prior ensembles of states at the input. In the case of the pure loss channel the result is given by the Gordon-Holevo capacity and equal to \cite{Gordon1962, Holevo2001, Giovannetti2004}
\begin{equation}\label{eq:holevo_ins}
     C_{\textrm{GH}}(\bar{n})=g(\tau_{\textrm{tot}}\bar{n}),
\end{equation}
where the function $g(x)=(x+1)\log_2(x+1)-x\log_2 x$. This formula can be further generalized to the case of general phase-insensitive Gaussian bosonic channels \cite{Giovannetti2014}. However, since the quantum channel implied by PSA is phase-sensitive, standard expressions for the capacity do not apply in this case. Instead, one rather needs to perform quite cumbersome calculations \cite{Schafer2016} resulting in 
\begin{equation}\label{eq:holevo_ph}
    C_{\textrm{GH}}(\bar{n})=g(\bar{M}_{\textrm{out}})-g(M_{\textrm{out}}),
\end{equation}
where $\bar{M}_{\textrm{out}}$ ad $M_{\textrm{out}}$ are quantities implicitly depending on the properties of the link which we discuss in the Appendix \ref{app:holevo}. Note that in the case of the pure loss channel, when the total transmission $\tau_{\textrm{tot}}=e^{-\alpha L}$ is small, both classical and quantum bounds predict capacity scaling as $C\sim e^{-\alpha L}$, which prevents communication on large distances.

A crucial difference between \eqnref{eq:holevo_ph} and \eqnref{eq:double} and \eqnref{eq:single} is that the quantum mechanical bound does not depend solely on the SNR. One needs to separately consider signal and noise in both quadratures in order to find the Gordon-Holevo capacity. For large noise, however, one may simplify the Gordon-Holevo bound. To see this, note that the function $g(x)$ for large $x$ can be approximated as
\begin{equation}
    g(x)\approx \log_2(1+x)+O\left(\frac{1}{x}\right).
\end{equation}
For large phase sensitive noise, i.e. $N^Q\gg 1$, one can approximate \eqnref{eq:Mout} and \eqnref{eq:Mbout} in Appendix \ref{app:holevo} by
\begin{equation}
    \bar{M}_{\textrm{out}}\approx y\sqrt{1+\textrm{SNR}^Q}-\frac{1}{2},\quad M_{\textrm{out}}\approx y-\frac{1}{2},
\end{equation}
where $y=\sqrt{N^QN^I}$. Plugging these values into \eqnref{eq:holevo_ph} gives
\begin{equation}
     C_{\textrm{GH}}(\bar{n})\approx \frac{1}{2}\log_2\left(1+\textrm{SNR}^Q\right),
\end{equation}
which is exactly \eqnref{eq:single}. Therefore, in the regime of large noise quantum effects are not relevant and the Gordon-Holevo capacity can be attained by the homodyne measurement irrespectively of the signal strength \cite{Jarzyna2019a}.

\section{Capacity of the PSA link}

The capacity of the PSA channel is given by either \eqnref{eq:single} if one insists on quadrature detection or \eqnref{eq:holevo_ph} if general quantum measurements are allowed. The signal and noise variances entering these formulas are given by \eqnref{eq:snr1}-(\ref{eq:snr2}). In order to find the ultimate bound on the capacity in both scenarios one needs to perform an optimization over the ensemble of input states and locations and gains of amplifiers. Assuming Gaussian input states, the first task can be done by simply optimizing the input signal and noise variances. In particular, the coherent state ensemble, which is most readily available, can be considered by just taking $N^{Q/I}_{\textrm{in}}=1/2$. On the other hand, optimization over the amplifiers has to be somehow constrained in order to avoid infinite gains of PSAs which would require large power at each amplifier. We will consider two types of constraint that have a simple physical interpretation: the amplitude restoration and total power restoration regimes.

\begin{figure}
	\centering
	\includegraphics[width=1.\columnwidth]{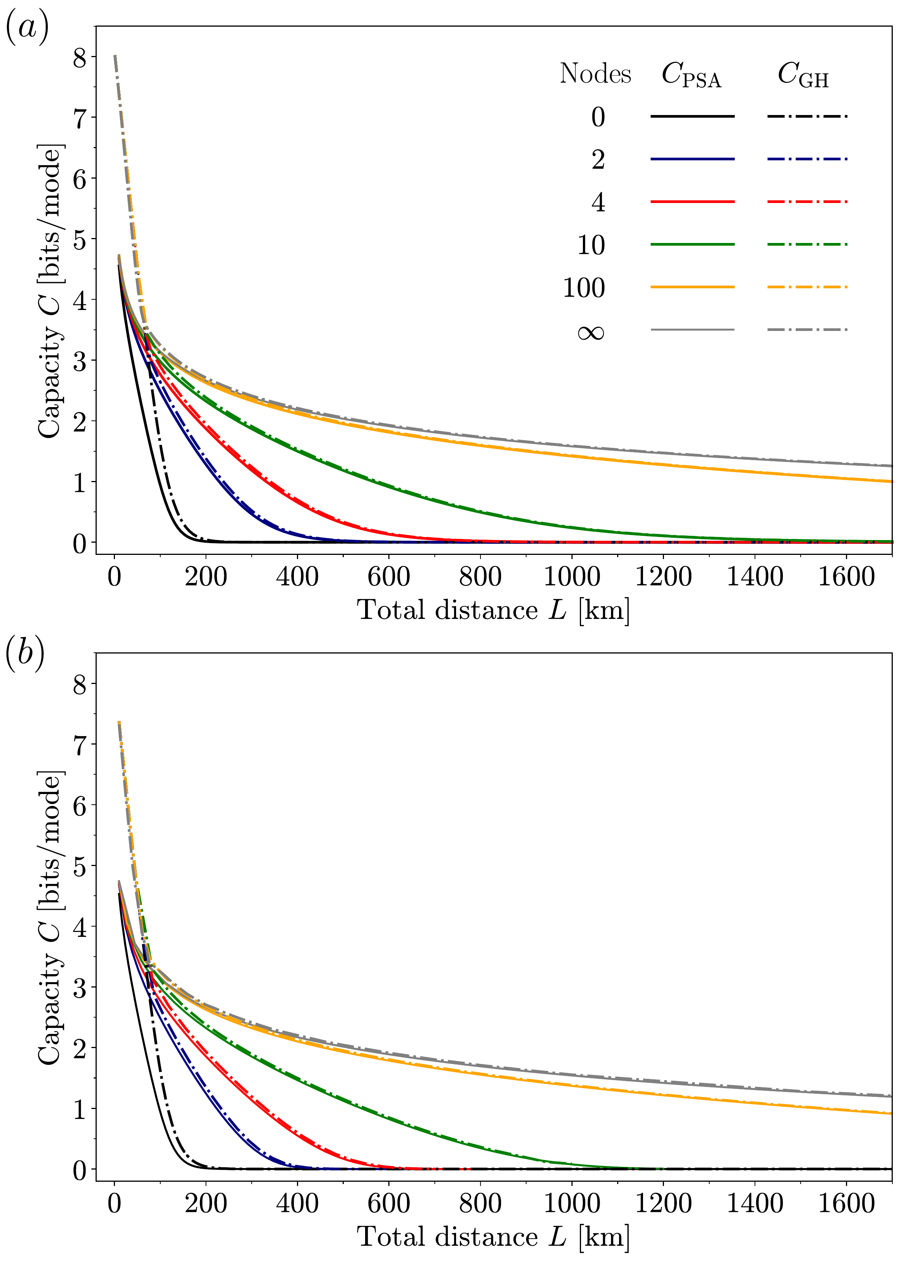}
	\caption{Capacity attainable with homodyne detection $C_\textrm{PSA}$ and the Gordon-Holevo bound $C_{\textrm{GH}}$ in the (a) amplitude and (b) total power restoration regimes as a function of the link length for $\alpha=0.05$ and $\bar{n}=100$.}
\label{fig:amplitude}
\end{figure}

In the first scenario, the amplitude restoration regime, each PSA restores the amplitude of the $Q$ quadrature to its initial value. In terms of gains and losses in each span this requirement translates into equality constraint $G_i=1/\tau_i$. The signal and noise Eqs.~(\ref{eq:var_q}),(\ref{eq:var_p}) therefore read
\begin{IEEEeqnarray}{rClrCl}
S^Q_i&=&S^{Q}_{\textrm{in}},\quad& N^Q_i&=&N^Q_{i-1}+\frac{1-\tau_i}{2\tau_i},\\
S^{I}_{i}&=&\tau_i^2S^{I}_{i-1},\quad& N^{I}_{i}&=&\tau_i^2N^I_{i-1}+\frac{\tau_i(1-\tau_i)}{2}.
\end{IEEEeqnarray}
Note that even though the signal in the $Q$ quadrature remains constant, the noise increases with each subsequent span of the link.  This is because not only the signal is amplified at each PSA but also the noise. A consequence of this fact is the growth of the total power traveling through the link with each passed PSA.

The constraints in the amplitude restoration regime can be easily formulated mathematically, but as argued above, in general they may lead to large total power in the fiber. This may be problematic since if the combined intensity of signal and noise becomes too strong, various non-linear effects begin to play a role in the propagation, which decreases the communication capacity \cite{Essiambre2010}. The workaround to avoid this issue is provided by the total power restoration regime in which it is the total power that cannot exceed its initial value at any point in the link. The relevant constraints read
\begin{equation}\label{eq:const}
    S^Q_i+S^I_i+N^Q_i+N^I_i\leq 2\bar{n}+1,
\end{equation}
for every amplifier $i$, which in terms of PSA gains is given by a set of complicated nonlinear inequalities. In general, the constraints in the total power restoration regime are stricter than in the amplitude restoration scenario, leading to a weaker output signal.

We performed numerical optimization over PSA gains and their locations in both approaches mentioned above. We chose the average number of photons $\bar{n}=100$ which corresponds to $\textrm{SNR}^{Q}=26 \textrm{dB}$ at the input. It is seen in Fig.~\ref{fig:amplitude} that the Gordon-Holevo capacity $C_{\textrm{GH}}$ can be improved by using PSA for distances $L\gtrsim 70 \textrm{km}$ for the typical value of the fiber attenuation coefficient $\alpha=0.05\textrm{km}^{-1}$. The advantage grows with the number of amplifiers and saturates at the curve representing the distributed amplification case, $R\to\infty$, which is discussed in detail in the next section. On the other hand, it is seen that the quantum advantage from using general POVMs is small when PSAs are applied and becomes negligible when the length of the link becomes considerable. Therefore, in this case one can attain the quantum bound using just the homodyne detection, as discussed in the previous section. In the regime of small distances $L\lesssim 70 \textrm{km}$ it is highly beneficial to utilize general quantum measurements, as the Gordon-Holevo bound is much larger than the Shannon-Hartley bound. However, in this limit, the capacity cannot be increased by signal amplification.

\section{Continuous amplification}
In general, finding the exact expressions for quadratures variances for optimal distribution of amplifiers is very complicated. However, one can derive simple formulas for the case of continuous amplification, when $R\to\infty$. In such a case, the discrete transmission of each span and PSA gains can be approximated as $\tau_i=1-\alpha\Delta l$ and $G_i=1+\gamma(l)\Delta l$, respectively. Note that although this approximation is performed assuming equal spans with length $\Delta l=L/(R+1)$, it remains valid in the general case. As discussed in \cite{Lukanowski2022} in the regime $\Delta l\to 0$, the propagation is approximated by
\begin{IEEEeqnarray}{rCl}\label{eq:snr_cont1}
        \frac{dS^Q}{dl}&=&\left(\gamma(l)-\alpha\right)S^Q,\\ \frac{dS^I}{dl}&=&-\left(\gamma(l)+\alpha\right)S^I, \\ \frac{dN^Q}{dl}&=&\left(\gamma(l)-\alpha\right)N^Q+\frac{\alpha}{2},\label{eq:snr_cont3}\\
    \frac{dN^I}{dl}&=&-\left(\gamma(l)+\alpha\right)N^I+\frac{\alpha}{2}.\label{eq:snr_cont4}
\end{IEEEeqnarray}
In case of amplitude restoration, one has $\gamma(l)=\alpha$, which, assuming coherent state modulation in the $Q$ quadrature, gives
\begin{gather}\label{eq:amp_rest}
        S^Q_{\textrm{out}}=2\bar{n},\quad S^I_{\textrm{out}}=0,\\ N^Q_{\textrm{out}}=\frac{1+\alpha L}{2},\quad N^I_{\textrm{out}}=\frac{1+e^{-2\alpha L}}{4}.
\end{gather}
Plugging these results into \eqnref{eq:single} gives in the large distance regime $\alpha L\gg 1$
\begin{equation}\label{eq:amp_asym}
    C\approx \frac{2\bar{n}}{\alpha L\ln 2}.
\end{equation}
This is an exponential gain with respect to the $C\sim e^{-\alpha L}\bar{n}$ value obtained for the pure loss channel without PSA.

For total power restoration the situation is more complicated since the gain function $\gamma(l)$ is given implicitly by the constraint in \eqnref{eq:const}. Nevertheless, one can still solve \eqnref{eq:snr_cont1}-(\ref{eq:snr_cont4}) analytically. For coherent state modulation in the Q quadrature the optimal gain profile reads
\begin{equation}\label{eq:gain}
    \gamma(l)=\frac{2\alpha\bar{n}}{\sqrt{4\bar{n}^2+2\bar{n}(1-e^{-2\alpha l})}},
\end{equation}
and the optimal quadrature variances are given in Appendix \ref{app:exact}. The capacity attainable with homodyne detection can then be approximated to 
\begin{equation}\label{eq:cap_power}
    C=\frac{1}{2}\log_2\left(1+\frac{4\bar{n}e^{-\frac{\alpha L}{4\bar{n}+1}}}{4\bar{n}\left(1-e^{-\frac{\alpha L}{4\bar{n}+1}}\right)+1}\right).
\end{equation}
In the large distance regime the above formula can be further approximated as 
\begin{equation}\label{eq:pow_asym}
C\sim \frac{2\bar{n}}{(4\bar{n}+1)\ln 2}e^{-\frac{\alpha L}{4\bar{n}+1}}.
\end{equation}
This represents an improvement in the exponent by factors of $1/(4\bar{n}+1)$ and $1/4$ with respect to the situation without PSA and when the aplifiers are phase-insensitive, respectively \cite{Jarzyna2019a}. On the other hand, the exponential decay of the capacity is still present, even in the optimal case.

\begin{figure}
	\centering
	\includegraphics[width=1.0\columnwidth]{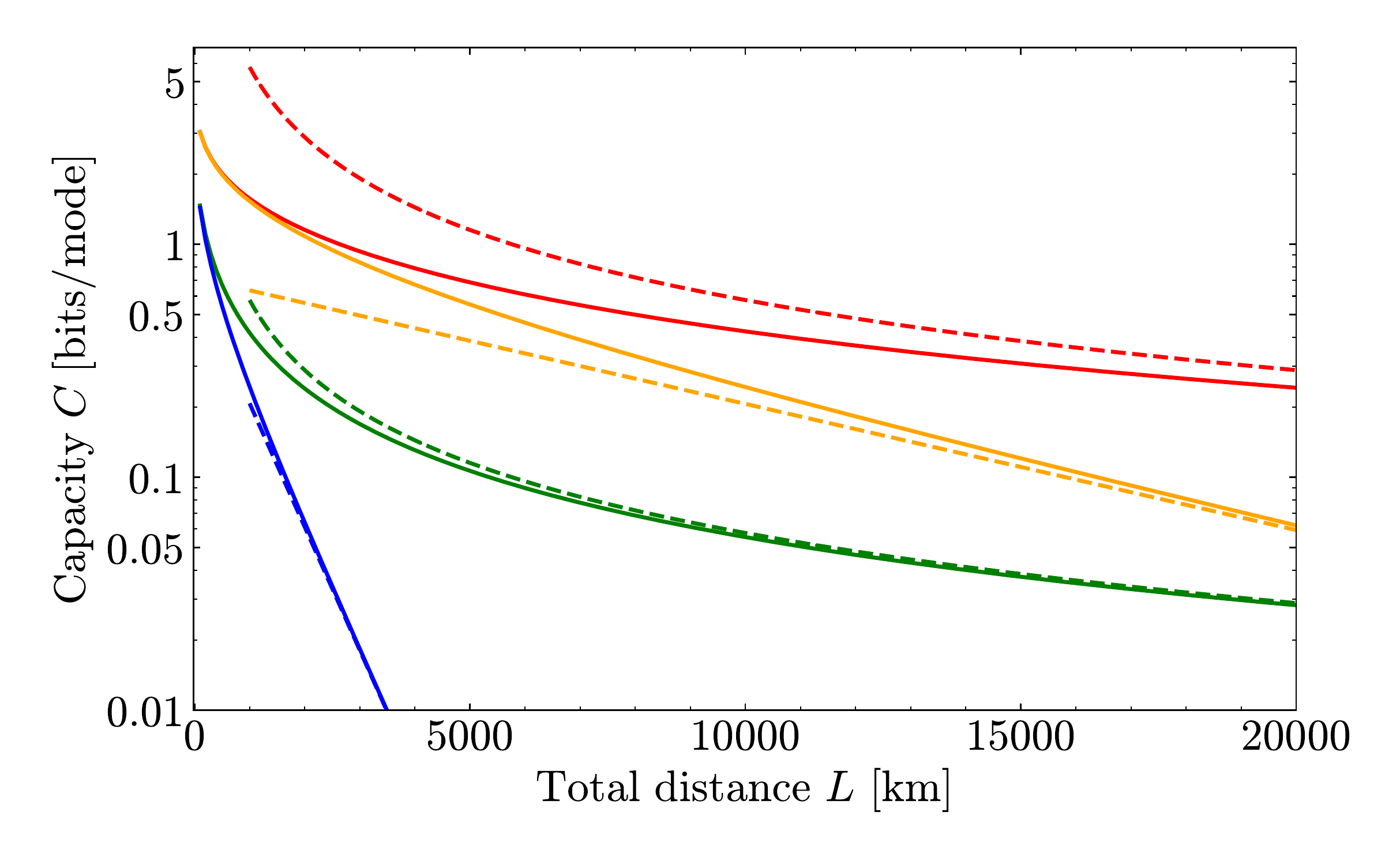}
	\caption{Optimized capacity as a function of the link length for $\alpha=0.05$. Power restoration regime: orange curve - $\bar{n}=100$, blue curve - $\bar{n}=10$; amplitude restoration regime: red curve - $\bar{n}=100$, green curve - $\bar{n}=10$. Dashed lines of respective colors represent corresponding asymptotic expressions, \eqnref{eq:amp_asym} for amplitude restoration and \eqnref{eq:pow_asym} for power restoration.}
\label{fig:asymptotic}
\end{figure}

In practical realizations, however, the average number of photons in the signal is large, $\bar{n}\gg 1$. This means that if $L\lesssim 4\bar{n}/\alpha$, the exponential term in \eqnref{eq:cap_power} in the denominator is nonegligible and for strong signals one can approximate it by $e^{-\alpha L/(4\bar{n}+1)}\approx 1-\frac{\alpha L}{4\bar{n}}$. Therefore, the capacity is rather equal to $C\sim \frac{1}{2}\log_2\frac{4\bar{n}}{\alpha  L}$ which is also the value obtained under the amplitude restoration constraint in this regime. One can intuitively understand this fact by inspecting the optimal gain profile in \eqnref{eq:gain}. Upon Taylor expansion for large $\bar{n}$ values one obtains $\gamma(l)\approx \alpha +O(1/\bar{n})$ which is very close to the gain in the amplitude restoration approach $\gamma_{\textrm{amp}}=\alpha$. Therefore, the capacities in these two instances should also be equal. This is seen in Fig.~\ref{fig:asymptotic} in which the attainable capacities in the amplitude and total power restoration regimes are equal for small and moderate distances and start to diverge for large link length, when they saturate respective asymptotic expressions.

\section{Conclusion}
In conclusion, we derived the ultimate quantum limits for the capacity of communication links with PSA. The advantage offered by general quantum mechanical measurement strategies against conventional protocols is small and becomes negligible in the long link regime. We showed that, depending on the constraints imposed on the amplification, one can consider two physically relevant regimes - amplitude and total power restoration. The asymptotic capacity values obtained in these two approaches differ significantly. Importantly, compared to pure loss and phase-insensitive amplification scenarios, the exponential decay rate of the capacity in the power restoration regime is improved by the factors of $1/(4\bar{n})$ and $1/4$ respectively, representing a huge advantage in the capacity. An even stronger enhancement can be found in the amplitude restoration regime in which the capacity scaling changes from exponential to inversely proportional in the link length. On the other hand, for strong signals, typically encountered in optical fiber links, both approaches turn out to be equivalent, offering exponential enhancement of the SNR. Crucially, our results suggest that quantum enhanced measurement strategies are beneficial only for short-haul links in which noise effects are not predominant.

\appendices

\section{Gordon-Holevo bound for the PSA channel}
\label{app:holevo}

In this section we outline the method to calculate the Gordon-Holevo capacity bound for phase-sensitive Gaussian channels described and proven in \cite{Schafer2016}. A general phase-sensitive single mode quantum Gaussian channel is specified by its action on the vector of the first moments $\mathbf{d}$ and the covariance matrix $\mathbf{\Sigma}$ of quadrature operators \cite{Weedbrook2011}. This can be described as
\begin{equation}
    \mathbf{d}_{\textrm{out}}=X\mathbf{d}_{\textrm{in}},\quad \mathbf{\Sigma}_{\textrm{out}}=X\mathbf{\Sigma}_{\textrm{in}}X^T+Y,
\end{equation}
where $X$ and $Y$ are real and symmetric matrices, satisfying certain conditions \cite{Weedbrook2011}. The von Neumann entropy of a general Gaussian state with a covariance matrix $\mathbf{\Sigma}$ is given by
\begin{equation}
    S(\rho)=g(M),\quad M=\sqrt{\det \mathbf{\Sigma}}-1/2.
\end{equation}
Assuming that the signal is modulated by displacing a certain Gaussian state with a given covariance matrix $\mathbf{\Sigma}_{\textrm{in}}$ in phase space  according to some Gaussian input distribution, the averaged input state in \eqnref{eq:holevo} is also Gaussian and is specified by the covariance matrix $\bar{\mathbf{\Sigma}}_{\textrm{in}}$. Therefore, the Gordon-Holevo capacity in such case is given by
\begin{equation}
    C_{\textrm{GH}}=g(\bar{M}_{\textrm{out}})-g(M_{\textrm{out}}),
\end{equation}
with
\begin{gather}
  \bar{M}_{\textrm{out}}=\sqrt{\det \bar{\mathbf{\Sigma}}_{\textrm{out}}}-1/2,\quad   \bar{\mathbf{\Sigma}}_{\textrm{out}}=X\bar{\mathbf{\Sigma}}_{\textrm{in}}X^T+Y,\\
  M_{\textrm{out}}=\sqrt{\det\mathbf{\Sigma}_{\textrm{out}}}-1/2,\quad   \mathbf{\Sigma}_{\textrm{out}}=X\mathbf{\Sigma}_{\textrm{in}}X^T+Y.
\end{gather}
This expression depends on the average input energy which reads
\begin{equation}
\nbar = \frac12 \left( \Tr\, \bar{\mathbf{\Sigma}}_{\textrm{in}} - 1 \right).
\end{equation}

One can show that for a given Gaussian channel one can come up with a so-called fiducial channel with the same capacity as the original one and diagonal matrices $X,\,Y$ \cite{Schafer2013}. The problem can then be described by just three parameters $\tau,\,y,\,\omega$ given by equations
\begin{equation}
\tau= \det X = \det \left(\begin{array}{cc}
x_{1} & 0\\
0 & x_{2}
\end{array}\right),\quad y=\sqrt{\det Y} 
\end{equation}
and
\begin{equation}
\label{eq:wenv}
Y= y \left(\begin{array}{cc}
\frac{x_1}{x_2} \omega^{-1} & 0\\
0 & \frac{x_2}{x_1} \omega
\end{array}\right).
\end{equation}
Note that the PSA channel link in consideration is already in the fiducial channel form, with $\tau=\tau_{\textrm{tot}}G_{\textrm{tot}}$, $y=\sqrt{N^QN^I}$, $\omega=\sqrt{\tau_{R+1}^2N^I/(\tau^2 N^Q)}$. These parameters calculated for a given channel allow one to specify a threshold energy
\begin{equation}
\nbar_\text{thr} = \frac{1}{2 \omega} \left( 1 + \frac{y}{|\tau|}|1-\omega^2| \right) - \frac12.
\end{equation}
If one wants to further optimize the Gordon-Holevo capacity formula in \eqnref{eq:holevo_ph} over ensembles of input states, the calculation proceeds based on whether the input energy is \emph{above} this threshold, $\nbar \geq \nbar_\text{thr}$, or \emph{below} it, $\nbar < \nbar_\text{thr}$.

\subsection{Above threshold}
We define
\begin{equation}
\wbin = \sqrt{\frac{|\tau| (2 \nbar + 1) + y(\omega^{-1}-\omega)}{|\tau| (2 \nbar + 1) - y \left(\omega^{-1}-\omega \right)}}.
\end{equation}
The optimal input covariance matrices read then
\begin{equation}
\mathbf{\Sigma}_{\text{in}}=\frac{1}{2}\left(\begin{array}{cc}
\frac{1}{ \omega} & 0\\
0 & \omega
\end{array}\right),\quad
\bar{\mathbf{\Sigma}}_{\text{in}}=\frac{2 \nbar +1}{\frac{1}{\wbin}+\wbin}\left(\begin{array}{cc}
\frac{1}{ \wbin} & 0\\
0 & \wbin
\end{array}\right),
\end{equation}
which corresponds to displacing in phase space a squeezed vacuum state according to a symmetric Gaussian distribution. The corresponding Gordon-Holevo capacity is given by
\begin{align}
 C_{\textrm{GH}}(\nbar) = &\;g\left[ |\tau|\left(\nbar + \frac12 \right) + \frac12 \left( y (\omega^{-1}+\omega )  - 1 \right) \right ] \nonumber\\ &- g\left[ \frac{|\tau|}{2} + y - \frac12\right ].
\end{align}

\subsection{Below threshold}
To calculate the classical capacity below threshold, one has to solve the following transcendental equation for $\omega_{\text{in}}$:
\begin{equation}
\label{eq:transc_below_thr}
\frac{g'\left(\Mbout \right)\ln2}{\wbout}\left(1-\wbout^{2}\right) = \frac{g' \left( \Mout \right) \ln 2 }{\wout} \left( \win^{2}-\wout^{2} \right)
\end{equation}
where $g'(x)$ is the derivative of $g(x)$ and we define
\begin{equation}
\label{eq:wout}
\wout=\sqrt{\frac{\tau\frac{\win}{2}+y \omega}{\tau\frac{1}{2 \win}+\frac{y}{\omega}}},
\end{equation}
\begin{equation}
\label{eq:wbout}
\wbout = \sqrt{\frac{\tau\left(2\nbar +1-\frac{1}{2\win}\right)+y \omega}{\tau\frac{1}{2 \win }+\frac{y}{\omega}}},
\end{equation}
\begin{equation}
\label{eq:Mout}
\Mout = -\frac{1}{2}+\sqrt{\left(\tau\frac{1}{2 \win }+\frac{y}{ \omega }\right)\left(\tau\frac{ \win }{2}+y \omega\right)},
\end{equation}
\begin{equation}
\label{eq:Mbout}
\Mbout=-\frac{1}{2}+\sqrt{\left(\tau\frac{1}{2 \win}+\frac{y}{ \omega }\right)\left(\tau\left(2 \nbar +1-\frac{1}{2 \win }\right)+y \omega\right)}.
\end{equation}
After finding $\win$ one then has
\begin{equation}
\wbin = \sqrt{2(2 \nbar +1) \win -1}.
\end{equation}
The optimal average input covariance matrix reads
\begin{equation}
\mathbf{\Sigma}_{\text{in}}=\frac{1}{2}\left(\begin{array}{cc}
\frac{1}{ \win} & 0\\
0 & \win
\end{array}\right),\quad
\bar{\mathbf{\Sigma}}_{\text{in}}=\frac{2 \nbar +1}{\frac{1}{\wbin}+\wbin}\left(\begin{array}{cc}
\frac{1}{ \wbin} & 0\\
0 & \wbin
\end{array}\right),
\end{equation}
which corresponds to displacing a squeezed vacuum state across a single quadrature. The Gordon-Holevo capacity below the threshold is finally given by
\begin{equation}
\label{eq:cap_bt}
 C_{\textrm{GH}}(\nbar) = g\left(\Mbout \right)-g\left( \Mout \right).
\end{equation}
Note that one can easily find what is the optimal capacity below the threshold assuming coherent state modulation by taking $\win=1$ in \eqnref{eq:Mout} and \eqnref{eq:Mbout} and plugging the result in \eqnref{eq:cap_bt}.

\section{Exact solution for distributed amplification}
\label{app:exact}
In order to find the exact quadrature variances with the constraint of total power \eqnref{eq:const} one can sum \eqnref{eq:snr_cont1}-(\ref{eq:snr_cont4}) resulting in
\begin{equation}
S^I+N^I=\frac{1}{\gamma(l)+\alpha}\left(\alpha+(\gamma(l)-\alpha)(S^Q+N^Q)\right),
\end{equation}
where we have used the fact that $dS^Q/dl+dS^I/dl+dN^Q/dl+dN^I/dl=0$ due to constant total power. Plugging this into \eqnref{eq:const} one obtains
\begin{equation}\label{eq:gamma}
    \gamma(l)=-\frac{2\alpha\bar{n}}{2\bar{n}+1-2(S^Q+N^Q)}.
\end{equation}
Let us now sum \eqnref{eq:snr_cont1} and \eqnref{eq:snr_cont3} and introduce $Z^Q=S^Q+N^Q$:
\begin{equation}
    \frac{dZ^Q}{dl}=-\alpha Z^Q\left(1+\frac{2\bar{n}}{2\bar{n}+1-2Z^Q}\right)+\frac{\alpha}{2}.
\end{equation}
The solution to this equation is given by
\begin{equation}
    Z^Q(l)=\frac{1}{2}\left[1+2\bar{n}+\sqrt{4\bar{n}^2+2\bar{n}-Ae^{-2\alpha l}}\right],
\end{equation}
where $A$ is a constant specified by initial conditions. For coherent state modulation in the $Q$ quadrature one has $A=2\bar{n}$. One can now plug this result in \eqnref{eq:gamma} and find the optimal gain profile
\begin{equation}
    \gamma(l)=\frac{2\alpha\bar{n}}{\sqrt{4\bar{n}^2+2\bar{n}-Ae^{-2\alpha l}}},
\end{equation}
which for $A=2\bar{n}$ gives \eqnref{eq:gain}. Using the derived value of $\gamma(l)$ one can now solve \eqnref{eq:snr_cont1}-(\ref{eq:snr_cont4}) resulting in
\begin{gather}\label{eq:sq1}
S^Q=2\bar{n}e^{-\alpha l}\left[\frac{1-\sqrt{\frac{1+2\bar{n}}{2\bar{n}}}}{1+\sqrt{\frac{1+2\bar{n}}{2\bar{n}}}}\times\frac{1+\sqrt{\frac{1+2\bar{n}}{2\bar{n}+1-e^{-2\alpha l}}}}{1-\sqrt{\frac{1+2\bar{n}}{2\bar{n}+1-e^{-2\alpha l}}}}\right]^{\sqrt{\frac{\bar{n}}{4\bar{n}+2}}},\\
    N^Q=Z^Q-S^Q,\\
    S^I=0,\\
    N^I=\frac{1}{2}\left[1+2\bar{n}-\sqrt{4\bar{n}^2+2\bar{n}-2\bar{n}e^{-2\alpha l}}\right],\label{eq:ni1}
\end{gather}
where we have taken $A=2\bar{n}$. By assuming large distances and then large power one can approximate the expressions for $S^Q$ and $N^Q$ by
\begin{gather}
S^Q=2\bar{n}e^{-\frac{\alpha L}{4\bar{n}+1}},\\
    N^Q=2\bar{n}(1-e^{-\frac{\alpha L}{4\bar{n}+1}})+\frac{1}{2},
\end{gather}
which results in capacity given in \eqnref{eq:cap_power}. On the other hand, by expanding  \eqnref{eq:sq1}-(\ref{eq:ni1}) for large input power $\bar{n}\gg 1$ these euqations can be approximated by
\begin{gather}
    S^Q\approx 2\bar{n}-\frac{\alpha L}{2}+\frac{1-e^{-2\alpha l}}{4},\\
    N^Q\approx \frac{1+\alpha L}{2},\\
    S^I=0,\\
    N^I\approx \frac{1+e^{-2\alpha l}}{4}.
\end{gather}
Note that this makes sense only when $\bar{n}\ll \alpha L/4$.
\section*{Acknowledgment}

The authors would like to thank Jochen Schr\"oder and Ren\'e-Jean Essiambre for insightful discussions. This work was supported by the Foundation for Polish Science under the ”Quantum Optical Technologies” project carried out within the International Research Agendas programme co-financed by the European Union under the European Regional Development Fund.

\ifCLASSOPTIONcaptionsoff
  \newpage
\fi

\bibliographystyle{IEEEtran}
\bibliography{IEEEabrv,pnr_channel}

\end{document}